\RequirePackage[l2tabu, orthodox]{nag}
\RequirePackage{snapshot}

\documentclass[10pt,twocolumn]{extarticle}
\makeatletter
\if@twocolumn
  \usepackage[dvips,letterpaper,top=0.5in, bottom=0.5in, left=0.75in, right=0.5in,includefoot,heightrounded]{geometry}
\else
  \usepackage[dvips,letterpaper,margin=1in,includefoot,heightrounded]{geometry}
\fi

\usepackage{srcltx}

\usepackage{abstract}

\usepackage{graphicx}

\usepackage{setspace}
\usepackage{url}

\usepackage[cmex10]{amsmath}
\usepackage[font=footnotesize,caption=false]{subfig}
\usepackage{color}
\usepackage[usenames,dvipsnames,svgnames,table]{xcolor}
\usepackage{array}
\usepackage{multirow}
\usepackage{textcomp}
\usepackage[para]{threeparttable}
\usepackage{rotfloat}

\usepackage{flushend}
\usepackage{picins}
\usepackage{wrapfig}

\usepackage{amssymb}

\usepackage[tiny]{titlesec}
\titleformat{\section}[hang]{\filright\scshape}{}{0em}{\thesection\quad}
\titleformat{\subsection}[hang]{\filright\scshape}{}{0em}{\thesubsection\quad}
\titleformat{\subsubsection}[hang]{\filright\scshape}{}{0em}{\thesubsubsection\quad}
\titleformat{name=\section,numberless}[hang]
{\filright\scshape}{}{0em}{}
\titleformat{name=\subsection,numberless}[hang]
{\filright\scshape}{}{0em}{}
\titleformat{name=\subsubsection,numberless}[hang]
{\filright\scshape}{}{0em}{}

\makeatletter

\newcommand{\printtitle}{%
\makeatletter
\if@twocolumn

\twocolumn[%
  \maketitle
  \begin{onecolabstract}
    \myabstract
  \end{onecolabstract}
  \begin{center}
    \small
    \textbf{Keywords}
    \\\medskip
    \mykeywords
  \end{center}
  \bigskip
]
\saythanks
\else
  \maketitle
  \begin{onecolabstract}
    \myabstract
  \end{onecolabstract}
  \begin{center}
    \small
    \textbf{Keywords}
    \\\medskip
    \mykeywords
  \end{center}
  \bigskip
  \onehalfspacing
\fi
\makeatother
}

\title{VLSI Computational
Architectures for the Arithmetic Cosine Transform}

\author{%
Nilanka~Rajapaksha
\quad
Arjuna~Madanayake%
\thanks{N. Rajapaksha and A. Madanayake are with the Department
          of Electrical and Computer Engineering, The University of Akron,
          Akron, OH, USA.
          E-mail: \{ntr3,arjuna\}@uakron.edu}
\quad
Renato~J.~Cintra%
\thanks{R. J. Cintra is with the Signal Processing Group,
          Departamento de Estat\'istica,
          Universidade Federal de Pernambuco,
          Recife, PE, Brazil.
          E-mail: rjdsc@ieee.org}
\\
Jithra~Adikari%
\thanks{J.~Adikari is with the Elliptic
          Technologies Inc., Ottawa, ON, Canada.
          E-mail: jithra.adikari@gmail.com}
\quad
Vassil~S.~Dimitrov%
\thanks{V. S. Dimitrov are with the
          Department of Electrical and Computer Engineering, University of Calgary,
          Calgary, AB, Canada.
          E-mail: vdvsd103@gmail.com}
}

\date{}

\newcommand{\myabstract}{%
The discrete cosine transform (DCT) is a widely-used and
important signal processing tool employed in a plethora of applications.
Typical fast algorithms for nearly-exact computation of DCT
require floating point arithmetic,
are multiplier intensive,
and accumulate round-off errors.
Recently proposed fast algorithm arithmetic
cosine transform (ACT) calculates
the DCT exactly using only additions and integer constant multiplications,
with very low
area complexity, for null mean input sequences.
The ACT can also be computed non-exactly for any input sequence,
with low area complexity and low power consumption,
utilizing the novel architecture described.
However, as a trade-off,
the ACT algorithm requires 10~non-uniformly
sampled data points to calculate the 8-point DCT.
This requirement can easily be satisfied for applications dealing with
spatial signals such as image sensors and biomedical sensor arrays,
by placing
sensor elements in a non-uniform grid.
In this work,
a hardware architecture for the computation of the
null mean ACT is proposed, followed by a novel
architectures that extend the ACT for non-null mean signals.
All circuits are physically implemented and tested using
the Xilinx XC6VLX240T FPGA device and synthesized for
45~nm TSMC standard-cell library for performance assessment.
}

\newcommand{\mykeywords}{%
Discrete cosine transform, Arithmetic cosine transform, fast algorithms, VLSI
}

\begin{document}

\printtitle

\section{Introduction}
\label{intro}

The discrete cosine transform (DCT) was first proposed by
Ahmed~\emph{et al.} in 1974
and published in \emph{IEEE Transactions on Computers}~\cite{ahmed1974discrete}.
It has since attracted much attention
in the computer engineering community~\cite{chakrabarti1990systolic,kamangar1982fast,kitajima1980symmetric,yu2001dct}.
In particular,
the 8-point \mbox{DCT} and its variants,
in the form of fast algorithms,
has been widely adopted
in several image and video coding standards~\cite{britanak2007cosine}
such as
JPEG,
\mbox{MPEG-1/2},
and
\mbox{H.261-5}~\cite{roma2007hybrid}.
Some applications which use image and video compression include
automatic surveillance~\cite{huei2009stereoscopic},
 geospatial remote sensing~\cite{geo1},
 traffic cameras~\cite{traffic1},
homeland security~\cite{surv2},
 satellite based imaging~\cite{satt1},
 unmanned aerial vehicles~\cite{uav1},
 automotive~\cite{marsi2007video},
 multimedia wireless sensor networks~\cite{wsn1},
 the solution of partial differential equations~\cite{proakis2007signal} etc.

A particular class of fast algorithms is constituted by
the arithmetic transforms.
An arithmetic transform is an algorithm for
low-complexity computation of a given trigonometric transform,
based on number-theoretical results.
A prominent example is the arithmetic Fourier transform (AFT)
proposed by Reed~\emph{et~al.}~\cite{reed1990mobius,reed1992vlsi}. The AFT allows multiplication-free calculation of Fourier coefficients using
number-theoretic methods and
non-uniformly sampled inputs.
A feature of the AFT is its suitability for
parallel implementation~\cite{reed1990mobius,reed1992vlsi}.

Recently,
an arithmetic transform method for the computation
of the DCT,
called the arithmetic cosine transform (ACT)
was proposed in~\cite{cintra2010act}.
The ACT can provide a multiplication-free framework
and leads to the
\emph{exact computation} of the DCT---provided
that the input signal has null-mean and is
non-uniformly sampled~\cite{cintra2010act}.
The computational gains of the ACT
are only possible when its prescribed non-uniformly sampled data is available.

Classically the required non-uniform samples
are derived by means of interpolation over
uniformly sampled data~\cite{cintra2010act}.
Such interpolation implies a computational overhead.
Another aspect
of the ACT
is that,
for arbitrary input signal,
it requires the computation of
the input signal mean value~\cite{cintra2010act}.
Usually,
the mean value is computed from uniformly sampled data~\cite{cintra2010act}.
In fact,
this dependence on uniformly sampled data
has been precluding
the implementation of the ACT based exclusively
on non-uniformly sampled data.

On the other hand,
the requirement for non-uniform samples can be
satisfied when
spatial input signals are considered.
In spatial signal processing,
non-uniformly sampled signals can be directly obtained,
without interpolation using a non-uniform placement of
sensors~\cite{tan2007cmos,roma2007hybrid}.
This motivates the search for architectures
which could solely rely on non-uniformly sampled inputs.

In this paper we address two main problems:
(i)~the proposition of a method to obtain the mean value of a given input
signal from its non-uniform samples as prescribed by the ACT
and
(ii)~the introduction of efficient
architectures for calculation of the 8-point DCT
based on the ACT, operating on non-uniformly sampled data \emph{only}.
This leads to designs with low computational complexity.
  Having ACT architectures that compute \mbox{1-D} DCT
  can be utilized as a building block to
  implement such \mbox{2-D} DCT architectures that
  take inputs from sensors placed on
  a non uniform grid.

Two architectures based on the ACT are sought,
being referred as Architectures~I and~II.
Architecture~I provides the hardware implementation of the ACT algorithm
proposed in \cite{cintra2010act}, and calculates the DCT with
exact precision for null mean 8-point sequences.
The proposed Architecture~I is designed to require
only
additions
and
multiplications by integers.
Thus,
no source of intrinsic computation error is present,
such as rounding-off and truncation.
Therefore,
area consuming hardware multipliers are not necessary.
We propose Architecture~II that implements the novel modified ACT algorithm
for DCT calculation of arbitrary, non-null-mean input signals, using 11 hardware
multiplications.
Both architectures require only non-uniformly sampled
inputs.

This paper unfolds as follows.
In Section~\ref{section-act},
the fundamental mathematical operations of the ACT
are briefly described.
Section~\ref{section-mean-matrix}
details how to compute the mean value from
non-uniformly sampled data and
provides a matrix formalism for the 8-point ACT.
In Section~\ref{section-vlsi}
the proposed architectures are detailed.
Section~\ref{section-results}
brings the implementation results
as well as comparisons with competing structures.
Conclusions and final remarks are furnished
in Section~\ref{section-conclusions}.

\section{The arithmetic cosine transform}
\label{section-act}

The usual input sequence to the DCT can be considered as
uniform samples of a continuous input signal~$v(t)$.
This results in an $N$-point column vector
$\mathbf{v} = \{ v_n \}_{n=0}^{N-1}$
which has its DCT
denoted by the $N$-point column vector~$\mathbf{V} = \{ V_k \}_{k=0}^{N-1}$.
To calculate~$\mathbf{V}$,
the ACT algorithm requires
non-uniformly sampled points of the continuous
input signal~$v(t)$~\cite{cintra2010act}.
These points are given by
\begin{align*}
r =
\frac{2mN}{k}
-
\frac{1}{2},
\end{align*}
where
$k=1,2,\ldots,N-1$,
and
$m=0,1,\ldots,k-1$~\cite{cintra2010act}.
We can define the set $R$ as:
\begin{align}
\label{eq.set_r}
R=\left\{\textrm{Set of all values of }r\right\}
\end{align}
It is important to
notice that the values of~$r$ are not necessarily integer.
In fact,
they are expected to be fractional.

If the signal of interest has zero mean,
then the ACT algorithm can be used to calculate
the DCT coefficients as follows.
First,
let the ACT averages $S_k$, $k=1,2,\ldots,N-1$,
of the non-uniform
sampled inputs be defined as~\cite{cintra2010act}:
\begin{align}
\label{eq.act-average}
S_k\triangleq
\frac{1}{k}
\sum_{m=0}^{k-1}
v_{2m
\frac{N}{k}-\frac{1}{2}},
\quad
k=1,2,\ldots N-1
.
\end{align}
The ACT averages can be employed to computed DCT coefficients
according to~\cite{cintra2010act}:
\begin{align}
\label{eq.act}
V_k
=
\sqrt{\frac{N}{2}}
\sum_{l=1}^{\left \lfloor\frac{N-1}{k} \right\rfloor}
\mu(l)
\cdot
S_{kl},
\quad
k=1,2,\ldots, N-1
,
\end{align}
where $\mu(\cdot)$ is the M\"obius
function~\cite{reed1990mobius,reed1992vlsi,cintra2010act}.
The derivation of the ACT~\cite{cintra2010act} utilizes
the M\"obius inversion formula.
Because the M\"obius function values are limited to $\{-1,0,+1\}$,
\eqref{eq.act} results in no additional multiplicative complexity.

In practice,
input sequences are not always null mean,
therefore
a correction term is necessary to~\eqref{eq.act}.
In~\cite{cintra2010act} an expression
suitable
for the non-null mean signals is given as:
\begin{align}
\label{eq.act.mertens}
V_k
=
\sqrt{\frac{N}{2}}
\sum_{l=1}^{\left\lfloor \frac{N-1}{k} \right\rfloor}
\mu(l) \cdot  S_{kl}
-
\sqrt{\frac{N}{2}}
\bar{v}
\cdot
\operatorname{M}
\left(
\left\lfloor \frac{N-1}{k} \right\rfloor
\right)
,
\end{align}
where
$\operatorname{M}(n) \triangleq \sum_{m=1}^n \mu(m)$
is the Mertens function~\cite{cintra2010act}
and~$\bar{v}$ is the mean value of the uniformly sampled input sequence.

\section{Proposed Algorithm with Only Non-uniformly Sampled Inputs}
\label{section-mean-matrix}

\subsection{Mean Value Calculation}

Although~\eqref{eq.act.mertens} leads to
the DCT coefficients of non-null mean input signals,
it requires the knowledge of quantity~$\bar{v}$,
which could be calculated straightforwardly from the~$N$ uniform samples
in~$\mathbf{v}$.
Since uniform samples are not available,
$\bar{v}$ should be directly calculated
from non-uniform samples.

The non-uniform samples are related to
the uniform samples according to
the interpolation scheme
given by~\cite{cintra2010act}:
\begin{align}
\label{eq.interpolation}
v_r
&=
\sum_{n=0}^{N-1}
w_n(r)
\cdot
v_n
,
\quad
r\in R
,
\end{align}
where $w_n(r)$ is the interpolation weight function
expressed by
\begin{align*}
w_n(r)
=
\frac{1}{2N}
\bigg[
&\operatorname{D}_{N-1}\left( \frac{\pi}{N} (n+r+1)\right)
+
\\
&
\operatorname{D}_{N-1}\left( \frac{\pi}{N} (n-r)\right)
\bigg],
\quad
n=0,1,\ldots,N-1
,
\end{align*}
and
\begin{align*}
\operatorname{D}_N(x) = \frac{\sin((N+1/2)x)}{\sin(x/2)}
\end{align*}
denotes the Dirichlet kernel~\cite[p.~312]{krantz2005real}.
Here, the set $R$ is defined in~\eqref{eq.set_r}.
More compactly,
\eqref{eq.interpolation}
can be put in matrix form.
Indeed,
we can write
$\mathbf{v_r} = \mathbf{W} \cdot \mathbf{v}$,
where
$\mathbf{v_r} = [ v_r ]_{r\in R}$
is a column vector containing the required non-uniform samples,
and $\mathbf{W} = \{ w_n(r) \}$, $n=0,1,\ldots,N-1$, $r\in R$,
is the implied interpolation matrix.

For the particular case of the 8-point ACT,
the following~10~non-uniform sampling instants
are required~\cite{cintra2010act}:
\begin{align}
\label{eq:act-sampling}
r \in R =
\left\{
-\frac{1}{2},
\frac{25}{14},
\frac{13}{6},
\frac{27}{10},
\frac{7}{2},
\frac{57}{14},
\frac{29}{6},
\frac{59}{10},
\frac{89}{14},
\frac{15}{2}
\right\}
.
\end{align}
Moreover,
matrix $\mathbf{W}$ is found to possess full column rank.
Thus,
its Moore-Penrose pseudo-inverse $\mathbf{W}^+$
is the left inverse of~$\mathbf{W}$~\cite[p.~93]{ipsen2009numerical}.
Therefore,
we obtain
\begin{align*}
\mathbf{v}
=
\mathbf{W}^+
\cdot
\mathbf{v_r}
.
\end{align*}
Consequently,
the mean value of $\mathbf{v}$ can be
determined exclusively from
the non-uniform samples,
according to:
\begin{align}
\label{equation-mean-value}
\bar{v}
=
\frac{1}{8}
\mathbf{w}
\cdot
\mathbf{v_r}
,
\end{align}
where $\mathbf{w}$ is the 8-point vector
of the sums of each column of~$\mathbf{W}^+$.
Scaled vector $\mathbf{w}/8$ has constant elements given by:
\begin{align*}
\frac{1}{8}
\mathbf{w}^\top
=
\begin{bmatrix}
\phantom{-}0.131763492716950 \\
\phantom{-} 0.498388117552161 \\
-0.313306526814540 \\
\phantom{-} 0.018837637958148 \\
\phantom{-} 0.389746948996966 \\
-0.178465262210960 \\
\phantom{-} 0.166302458810496 \\
\phantom{-} 0.269801852271683 \\
-0.131541981375149 \\
\phantom{-} 0.148473262094246
\end{bmatrix}
,
\end{align*}
were the superscript ${}^\top$ denotes the transposition operation.

\subsection{Matrix Factorization of ACT}

In view of~\eqref{eq.act-average} and~\eqref{equation-mean-value},
\eqref{eq.act.mertens} can be interpreted as the sought relation
between~$\mathbf{V}$ and $\mathbf{v_r}$.
Thus,
we can consider a transformation matrix~$\mathbf{T}$
relating these two vectors.
Notice that~$\mathbf{T}$ is not a square matrix.
Since $k=1,2,\ldots,N-1$,
the size of~$\mathbf{T}$ is $(N-1) \times |R|$,
where $|R|$ is the number of elements of $R$.
This transformation matrix
returns all the DCT components,
except the zeroth one,
according to:
\begin{align*}
\begin{bmatrix}
V_1 & V_2 & \cdots & V_{N-1}
\end{bmatrix}^\top
=
\mathbf{T}
\cdot
\mathbf{v_r}.
\end{align*}
Notice that $V_0 = \sqrt{N} \cdot \bar{v}$.

For $N=8$,
matrix $\mathbf{T}$ has size 7$\times$10
and
admits the following matrix factorization:
\begin{align}
\label{equation-factorization}
\mathbf{T}
=
2
\cdot
\mathbf{Mo}
\cdot
\mathbf{D_1}
\cdot
\mathbf{S}
+
\mathbf{Me}
\cdot
\mathbf{W^+}
,
\end{align}
where
\begin{align*}
\mathbf{Mo}=\left[
    \begin{array}{rrrrrrr}
      1&-1&-1&0&-1&1&-1\\
      0&1&0&-1&0&-1&0\\
      0&0&1&0&0&-1&0\\
      0&0&0&1&0&0&0\\
      0&0&0&0&1&0&0\\
      0&0&0&0&0&1&0\\
      0&0&0&0&0&0&1
    \end{array}
  \right]
,
\end{align*}
\begin{align*}
\mathbf{D_1}
=
\begin{bmatrix}
1&0&0&0&0&0&0\\
0&\frac{1}{2}&0&0&0&0&0\\
0&0&\frac{1}{3}&0&0&0&0\\
0&0&0&\frac{1}{4}&0&0&0\\
0&0&0&0&\frac{1}{5}&0&0\\
0&0&0&0&0&\frac{1}{6}&0\\
0&0&0&0&0&0&\frac{1}{7}
\end{bmatrix}
,
\end{align*}
\begin{align*}
\mathbf{S}
=
\begin{bmatrix}
      1&0&0&0&0&0&0&0&0&0\\
      1&0&0&0&0&0&0&0&0&1\\
      1&0&0&0&0&0&2&0&0&0\\
      1&0&0&0&2&0&0&0&0&1\\
      1&0&0&2&0&0&0&2&0&0\\
      1&0&2&0&0&0&2&0&0&1\\
      1&2&0&0&0&2&0&0&2&0
\end{bmatrix}
,
\end{align*}
and
$\mathbf{Me}$ is the implied matrix by the Mertens function
in~\eqref{eq.act.mertens}.
This last matrix is furnished by
\begin{align*}
\mathbf{Me}
=
\begin{bmatrix}
\frac{1}{2}&0&0&0&0&0&0\\
0&\frac{1}{4}&0&0&0&0&0\\
0&0&0&0&0&0&0\\
0&0&0&-\frac{1}{4}&0&0&0\\
0&0&0&0&-\frac{1}{4}&0&0\\
0&0&0&0&0&-\frac{1}{4}&0\\
0&0&0&0&0&0&-\frac{1}{4}
\end{bmatrix}
\cdot
\mathbf{1}_7
,
\end{align*}
where
$\mathbf{1_7}$ is the 7$\times$7 matrix of ones.

In~\eqref{equation-factorization},
matrix~$\mathbf{D_1}$ and $\mathbf{S}$
are related to~\eqref{eq.act-average}.
Matrix~$\mathbf{Mo}$ contains the values
of the M\"obius functions as required in~\eqref{eq.act}.
The second term in the right-hand side of~\eqref{equation-factorization}
accounts for the Mertens functions and the mean value calculation
as required in~\eqref{eq.act.mertens} and~\eqref{equation-mean-value}.

\section{VLSI Architectures}
\label{section-vlsi}

In this section,
above discussed methods
are employed to
furnish
two novel
low complexity architectures,
which take only non-uniformly sampled inputs.
Integer multiplications, which are exact in nature,
are realized using simple shift-add structures.
The designs are fully pipelined by judicious insertion of
registers at internal nodes,
leading to low critical path delay at the
cost of latency.

\subsection{Architecture~I}

\begin{figure*}
\centerline{
\subfloat[Architecture~I]{
\includegraphics[width=11cm]{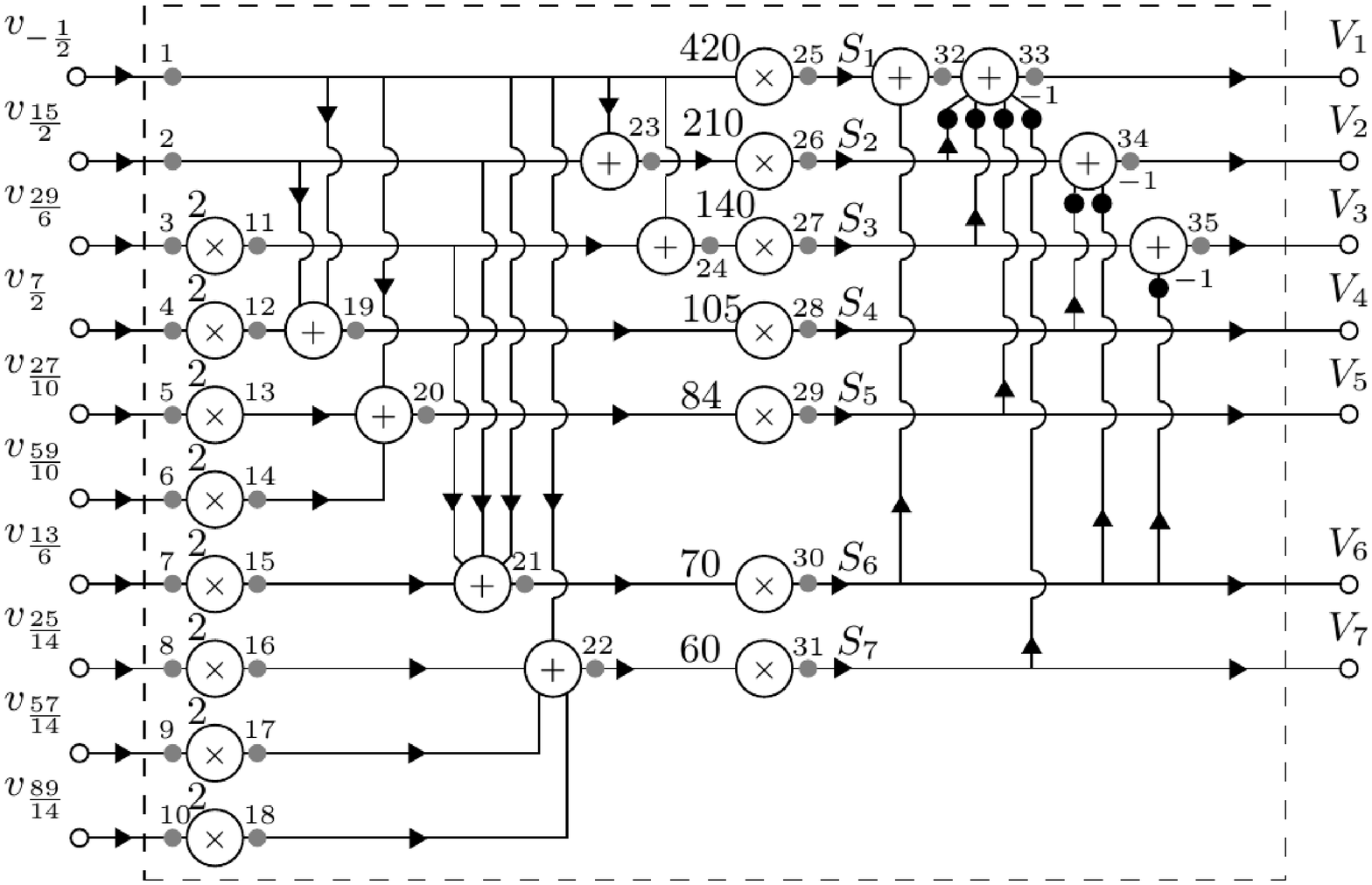}
\label{fig:null_mean_block}}
\hfil
\subfloat[Mean calculation block]{
\includegraphics[width=6.5cm]{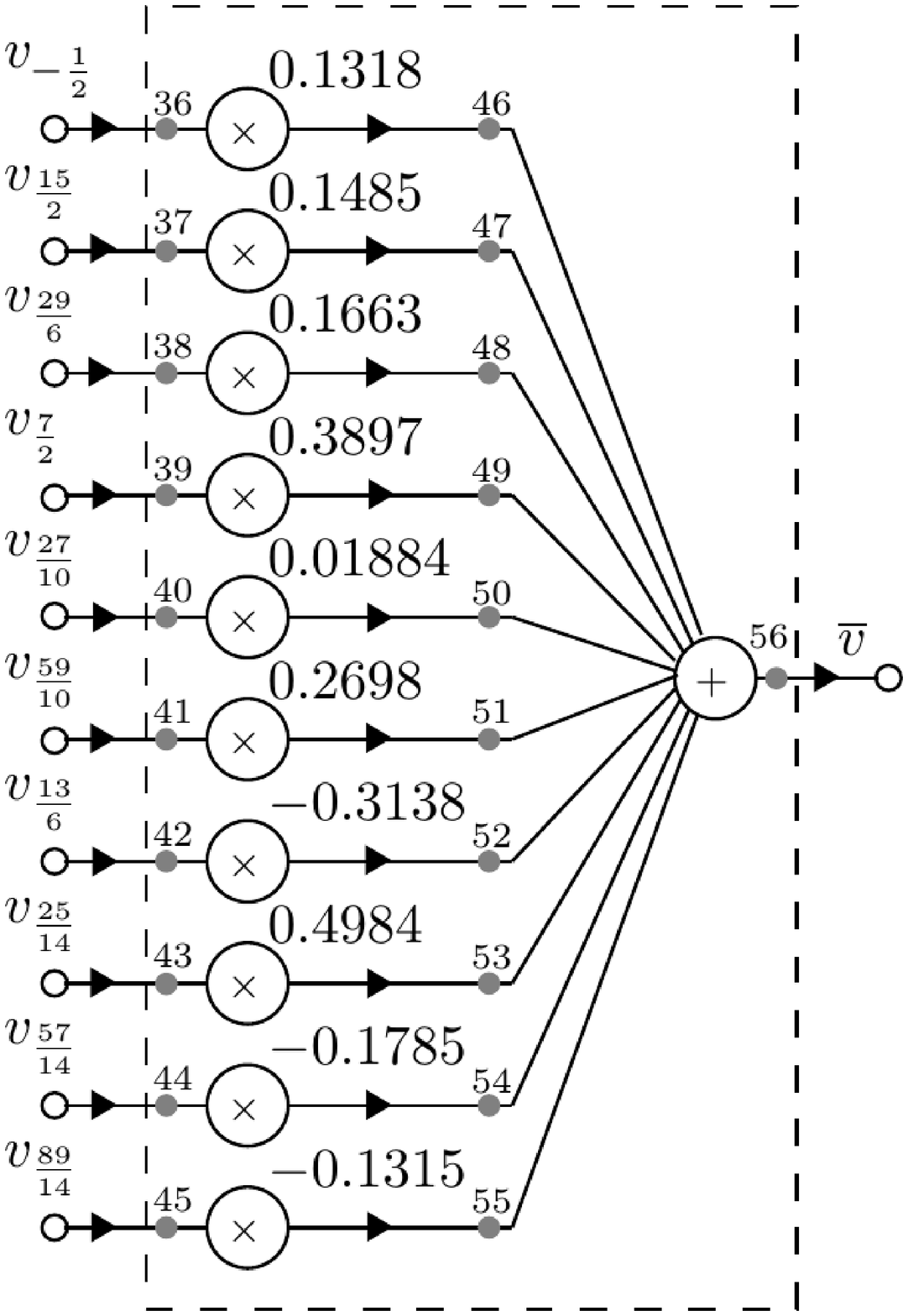}
\label{fig:mean_substraction}}
}
\caption{(a)~Null mean ACT and (b)~mean calculation block.}
\label{figure-arch-I-and-mean}
\end{figure*}

The ACT expressions for null mean signals
in~\eqref{eq.act-average} and~\eqref{eq.act}
can be implemented for $N=8$ as
shown in Fig.~\ref{figure-arch-I-and-mean}(a).
We refer to this design as Architecture~I.
The 8-point null mean ACT block admits the 10~non-uniformly sampled
inputs corresponding to~\eqref{eq:act-sampling}.
Constant multiplications by~two are
implemented as left-shift operations;
and the fractional constant multipliers
$1/2, 1/3, 1/4, \ldots$
are converted to integers by
scaling them by the least common multiple of their denominators:
420.
The integer constant multipliers can be implemented as
Booth encoded shift-and-add structures
making the architecture multiplier free, and the outputs of the block
are scaled by $420\cdot\sqrt{2/N}=210$.
This architecture is useful in applications
that have null mean input sequences,
and can be implemented with very low computational and area complexity.

\begin{figure*}
\centerline{
\subfloat[Mertens correction block]{
\includegraphics[width=6cm]{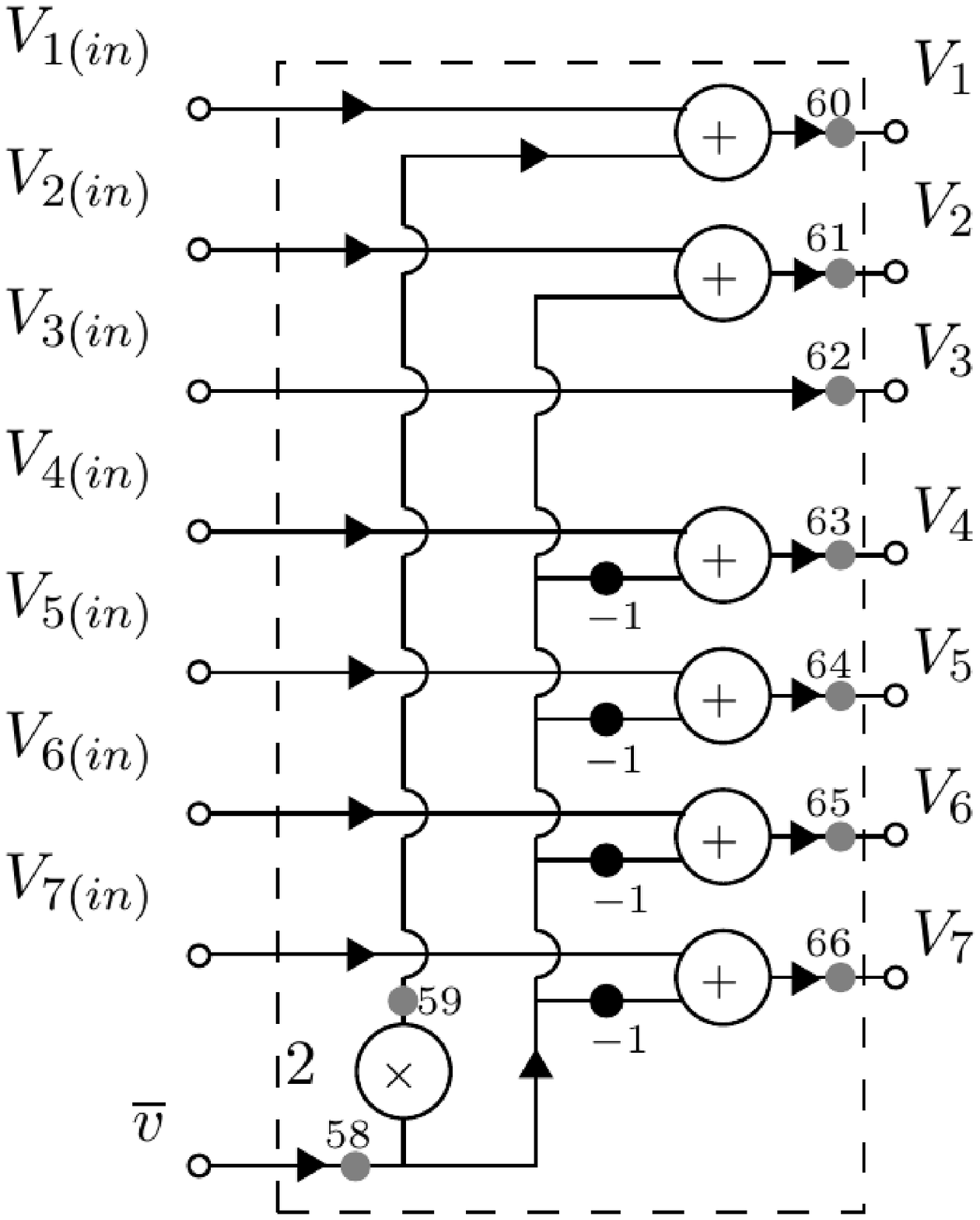}
\label{fig:mertens_correction}
}
\hfil
\subfloat[Non-null mean ACT]{
\includegraphics[width=10cm]{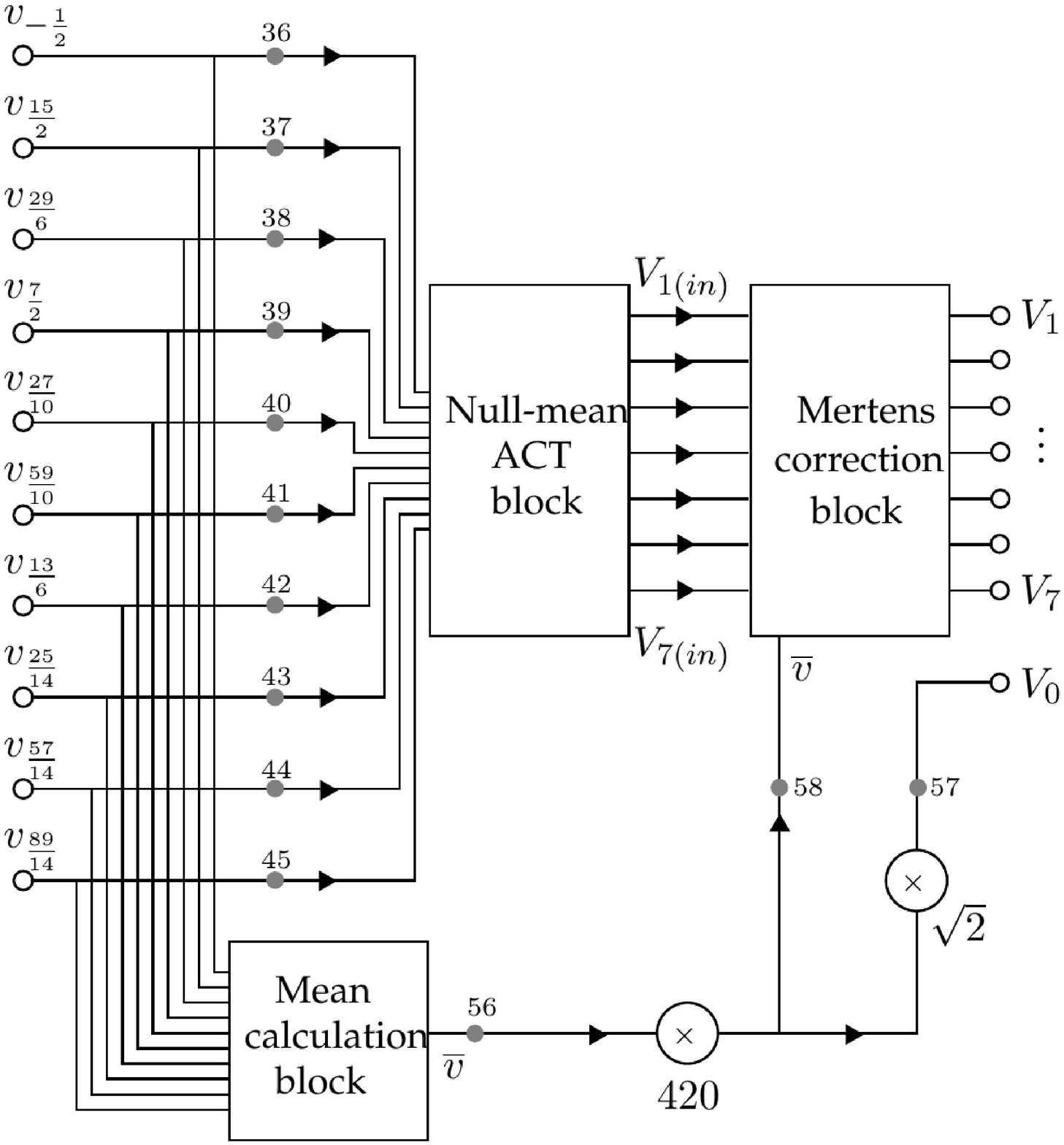}
\label{fig:mertens_correction2}}}
\caption{Architecture~II:
Non-null mean DCT calculation using the Mertens correction block.}
\label{figure-arch-II}
\end{figure*}

\subsection{Architecture~II}

The proposed method in Section~\ref{section-mean-matrix} for the computation of
$\bar{v}$ from the non-uniformly sampled 10-point signal
can be implemented
as shown in Fig.~\ref{figure-arch-I-and-mean}(b).
We will refer to it as the mean calculation block,
which computes~\eqref{equation-mean-value}.
The correction term associated to the Mertens function
required in~\eqref{eq.act.mertens}
is shown in Fig.~\ref{figure-arch-II}(a).
A combination of
(i)~this particular block,
(ii)~the Architecture~I block,
and
(iii)~the mean calculation block
yields
the
proposed Architecture~II as shown in
Fig.~\ref{figure-arch-II}(b).

Note that calculation of the DCT coefficients using the null mean ACT block
can also be achieved by
subtracting the mean~$\bar{v}$ from its inputs.
However,
Architecture~II has a lower
computational complexity when compared to
such alternative.
Computation complexity of both Architecture~I and Architecture~II are listed
in Table~\ref{tab:complex}
in terms of constant multipliers and two-input adders.
Integer constant multiplications are implemented as shift-and-add structures,
therefore are not counted as multipliers.
Note that the adder count also include the adders required for the Booth encoded structures.
\begin{table}
\centering
\caption{Computational complexity of proposed Architecture~I and
Architecture~II.}
\label{tab:complex}

\begin{tabular}{|>{\centering}m{3cm}|>{\centering}m{1.9cm}|>{\centering}m{1.9cm}|}
\cline{2-3}
\multicolumn{1}{>{\centering}m{3cm}|}{} & Architecture~I & Architecture~II\tabularnewline
\hline
Constant multipliers & 0 & 11\tabularnewline
Two-input Adders & 36 & 54\tabularnewline
\hline
\end{tabular}
\vskip-2ex
\end{table}

\section{Implementation and Results}
\label{section-results}

\subsection {FPGA Implementation}

We implemented both architectures described in
the previous section.
These architectures were tested on Xilinx \mbox{Virtex-6} XC6VLX240T FPGA using the stepped hardware co-simulation
feature in ML605 evaluation platform.
They were also fully pipelined to achieve the maximum throughput.
Word-length is $L$ at the inputs,
which are assumed to be in the range $[-1,1]$.
Throughout
the fixed point implementation
the word-length increases to avoid overflow.
Depending on the particular quantization point,
the actual allocated word-length is given by
$L + \Delta L$,
where the values of $\Delta L$ are listed
in Table~\ref{tab:fix_alloc}
for both proposed architectures.
The referred quantization points are shown
in Fig.~\ref{figure-arch-I-and-mean} and Fig.~\ref{figure-arch-II}.
The number of fractional bits are maintained constant throughout
the design and is equal to $L-1$.

Accuracy of the results from Architectures~I and~II
were tested with varying values of $L$ by using average percentage error and peak signal to noise
ratio~(PSNR) as figures of merit.
Adopted figures of merit employed
the DCT coefficients calculated from the
floating point implementation of the DCT
available in Matlab
as reference.
Results given in Table~\ref{tab:error}
are taken from the simulation of Architectures~I and II using
$10^4$ random input signals.
The reduction of the input word-length $L$
degrades the results furnished by the considered figures of merit.
However,
for small word-lengths,
the errors incurred are tolerable for most applications.

\begin{table}
\centering

\caption{Fixed point word-length increase $\Delta L$ at each
quantization point of the ACT signal flow graph.
Fixed point word-length is $L + \Delta L$}
\label{tab:fix_alloc}

\begin{tabular}{|>{\centering}m{2cm}|>{\centering}m{0.7cm}|>{\centering}m{2cm}|>{\centering}m{0.7cm}|}
\hline
\multicolumn{2}{|c|}{Architecture~I}  & \multicolumn{2}{c|}{Architecture~II}\tabularnewline
\hline
Points & $\Delta L$ & Points & $\Delta L$  \tabularnewline
\hline
1--10 &  0 &  36--55 &  0  \tabularnewline
11--18 &  2 &  56 &  1  \tabularnewline
19--22, 24 &  3 &  57, 59, 61, 62 &  13 \tabularnewline
23 &  1 &  58 &  11  \tabularnewline
25, 26, 31 &  10 &  60 &  14  \tabularnewline
27, 32, 34 &  12 &  63--66 &  12  \tabularnewline
\cline{3-4}
28--30 &  11 &  \multicolumn{2}{c}{}  \tabularnewline
33, 35 &  13 &  \multicolumn{2}{c}{} \tabularnewline
\cline{1-2}
\end{tabular}
\vskip-2ex
\end{table}

\begin{table*}
\centering

\caption{Average percentage error and
 average peak signal to noise ratio (PSNR) of ACT
implementations with fixed point input word-length $L$,
when tested with 10,000 input vectors}
\label{tab:error}

\begin{tabular}{|>{\centering}m{0.5cm}|>{\centering}m{2cm}|>{\centering}m{1.5cm}|>{\centering}m{2cm}|>{\centering}m{1.5cm}|}
\hline
\multirow{2}{0.5cm}{\centering $L$}  &
\multicolumn{2}{>{\centering}m{3.2cm}|}{Architecture~I} &
\multicolumn{2}{>{\centering}m{3.2cm}|}{Architecture~II}

\tabularnewline

\cline{2-5}

 & \% error & PSNR~(dB) & \% error & PSNR~(dB) \tabularnewline
\hline
 8  &   ${ 4.594\times 10^{-1}}$ &  50.3 &  ${ 2.262\times 10^{-1}}$ &  38.8 \tabularnewline
 12  &   ${ 1.977\times 10^{-2}}$ &  74.3 &  ${ 2.149\times 10^{-1}}$ &  63.0 \tabularnewline
 16  &   ${-1.840\times 10^{-3}}$ &  98.4 &  ${ -1.550\times 10^{-2}}$ &  87.1 \tabularnewline
 20  &   ${ 2.943\times 10^{-4}}$ &  122.4 &  ${ 2.565\times 10^{-3}}$ &  110.8 \tabularnewline
 24  &   ${ -1.001\times 10^{-5}}$ &  145.6 &  ${ 9.462\times 10^{-6}}$ &  135.4 \tabularnewline
 28  &   ${ 1.167\times 10^{-6}}$ &  170.6 &  ${ 3.137\times 10^{-6}}$ &  159.4 \tabularnewline
 32  & ${ -2.274\times 10^{-8}}$ &  194.7 &  ${ 3.207\times 10^{-7}}$ &  183.4 \tabularnewline
\hline
\end{tabular}
\end{table*}

Table~\ref{tab:resource} shows the resource utilization, power consumption and operational frequency
on the Xilinx Virtex-6 XC6VLX240T FPGA device
for input fixed point word-lengths ($L$)~8 and 12.
Information about the Xilinx FPGA resources that are
listed in Table~\ref{tab:resource} including
slices, slice FFs and 4-input look-up tables (LUTs)
can be found in the device datasheet.
Architecture~I is multiplier-free and
possesses the lower complexity,
but it is only suitable for null mean signals.
To remove the dependence of power consumption to operating frequency the
normalized power metric (dynamic power normalized to operating frequency)
is given in Table~\ref{tab:resource}.
The total power consumption in the FPGA is dominated by the static power
since both architectures only occupied roughly 1\% of the available area.

\begin{table*}
\caption{\label{tab:resource}Speed of operation resource utilization and power consumption of the XC6VLX240T FPGA device used for input fixed point word-lengths $L$ and for Architectures~I and II}
\centering
\begin{tabular}{|>{\centering}m{1.86cm}|>{\centering}m{1.5cm}|>{\centering}m{1.5cm}|>{\centering}m{0.7cm}|>{\centering}m{0.7cm}|>{\centering}m{1.5cm}|>{\centering}m{1.5cm}|>{\centering}m{2cm}|}
\hline
\centering Architecture tested &
\centering Fixed point word-length ($L$)&
\centering Slices &
\centering Slice FF &
\centering Slice LUTs &
\centering Dyn. power (W) &
\centering Op. freq. (MHz) &
\centering Norm. power (W/MHz)\tabularnewline
\hline
\multirow{2}{1.86cm}{\centering Architecture I} & 8 & 263 (1\%)& 930 (1\%) & 756 (1\%) & 1.37 & 500 & $2.74\times 10^{-3}$ \tabularnewline
 & 12 & 329 (1\%) & 1205 (1\%) & 1019 (1\%) & 1.16 & 333.33 & $3.49\times 10^{-3}$ \tabularnewline
\hline
\multirow{2}{1.86cm}{\centering Architecture II} & 8 & 443 (1\%) & 1276 (1\%) & 1386 (1\%) & 0.54 & 166.66 & $3.22\times 10^{-3}$ \tabularnewline
 & 12 & 495 (1\%) & 1678 (1\%) & 1639 (3\%) & 0.53 & 133.33 & $3.97\times 10^{-3}$ \tabularnewline
\hline
\end{tabular}
\end{table*}

\subsection {ASIC Synthesis Results}

The proposed architecture Architecture I and II are synthesized for application
specific integrated circuits (ASIC) using the
Cadence RTL Compiler for 45~nm technology.
The freePDK45 standard-cell library is used in synthesis
with optimization goal
set to maximize the speed.
Our synthesis was performed at operating voltage of 1.1~V.
The area,
power,
operational frequency,
and
normalized power metric (dynamic power normalized to operating frequency and square of the supply voltage)
for
the ASIC synthesis are presented in
Table~\ref{tab:resource2}.

\begin{table*}
\caption{Speed of operation, critical path delay, power consumption and area utilization in ASIC synthesis results for fixed point word-lengths $L$ for Architecture~I (45\,nm technology)}
\label{tab:resource2}
 \centering
 \begin{tabular}{|
>{\centering}m{1.97cm}|
>{\centering}m{0.9cm}|
>{\centering}m{1.2cm}|
>{\centering}m{0.7cm}|
>{\centering}m{1.1cm}|
>{\centering}m{1.0cm}|
>{\centering}m{1.0cm}|
>{\centering}m{1.20cm}|
>{\centering}m{1.0cm}|}
 \hline
 Architecture Synthesized &
 Fixed point word-length ($L$)&
 Area ($\mu\text{m}^2$) &
 Static power (mW) &
 Dynamic power (mW) &
 Total power (mW) &
 Op. Freq. (GHz) &
 Norm. Freq. (mW /GHz$\cdot$V$^2$)\tabularnewline
 \hline
 \multirow{2}{1.97cm}{\centering Architecture I} & 8 & 39007.27  & 0.27 & 67.31 & 67.59 & 1.11 & 50.12 \tabularnewline
 & 12 & 53961.52 & 0.37 & 90.32 & 90.70 & 1.11 & 67.25 \tabularnewline
 \hline
 \multirow{2}{1.97cm}{\centering Architecture II} & 8  & 65314.36 & 0.46 & 60.34 & 60.80 & 0.625 & 79.78 \tabularnewline
 & 12  & 96087.77  & 0.63 & 79.29 & 79.92 & 0.588 & 111.45 \tabularnewline
 \hline
 \end{tabular}
 \end{table*}

Table~\ref{tab:prev} shows the comparison of results between
proposed ACT Architectures I and II and other published 8-point DCT implementations.
Ideally,
a fair comparison requires all implementations
to be of the same process,
operating frequency, and supply voltage.
However,
the published literature contains varying technology and
operational conditions.
Hence in Table~\ref{tab:prev} a normalized power consumption value is given, where the power consumption is normalized
to the corresponding operational frequency and square of supply voltage.
From the normalized power consumption given in Table~\ref{tab:prev} it's apparent
that the proposed architectures consume lower power
than architectures in \cite{gong_2004},\cite{Gosh_2005} and \cite{livramento}.
We emphasize that the proposed
Architecture~I has the distinct advantage of having
\emph{exact} computation.
Thus approximate DCT methods
as suggested in~\cite{dct_approx1,dct_approx2}
were not taken into consideration
for comparison purposes.

\begin{table*}
  \centering
  \begin{threeparttable}[b]
  \caption{\label{tab:prev}
Comparison of the proposed implementation with published DCT implementations. Some implementations are \mbox{2-D}
    but since they are implemented with \mbox{1-D} DCT module with row column decomposition, results can be taken that can be compared with the proposed architectures.}

  \begin{tabular}{%
|>{\centering}m{2.3cm}
|>{\centering}m{1.3cm}
|>{\centering}m{1.3cm}
|>{\centering}m{1.3cm}
|>{\centering}m{1.3cm}
|>{\centering}>{\bfseries}m{1.3cm}
|>{\centering}>{\bfseries}m{1.3cm}
|>{\centering}>{\bfseries}m{1.3cm}
|>{\centering}>{\bfseries}m{1.3cm}|}

\cline{2-9}
    \multicolumn{1}{>{\centering}m{1cm}|}{} &
\multirow{2}{1.3cm}{\centering\scriptsize Gong \emph{et~al.} \cite{gong_2004}} &
\multirow{2}{1.3cm}{\centering\scriptsize Shams \emph{et~al.} \cite{Shams_2000}} &
\multirow{2}{1.3cm}{\centering\scriptsize Gosh \emph{et~al.} \cite{Gosh_2005}} &
\multirow{2}{1.3cm}{\centering\scriptsize Livramento \emph{et~al.}\\ \cite{livramento}} &
\multicolumn{4}{>{\centering}m{4.8cm}|}{\scriptsize Proposed architectures}
\tabularnewline
    \cline{6-9}
    \multicolumn{1}{>{\centering}m{0.9cm}|}{}& & & & & \scriptsize Arch.\\ I & \scriptsize Arch.\\ I & \scriptsize Arch.\\ II & \scriptsize Arch.\\ II
\tabularnewline
    \hline
    \scriptsize 1D/2D DCT & 2D but 1D results available & 1D & 2D but 1D results available & 2D but 1D results available & 1D & 1D & 1D & 1D \tabularnewline
    \scriptsize Replicated and measured results by authors & \scriptsize No & \scriptsize No & \scriptsize No & \scriptsize No &\scriptsize  Yes &\scriptsize  Yes &\scriptsize  Yes &\scriptsize  Yes\tabularnewline
    \scriptsize Precision & \scriptsize Non-exact & \scriptsize Non-exact & \scriptsize Non-exact & \scriptsize Non-exact & \scriptsize Exact & \scriptsize Exact & \scriptsize Non-exact & \scriptsize Non-exact \tabularnewline
    \scriptsize Method & \scriptsize Vector matrix DCT core & \scriptsize New distributed arithmetic based DCT & \scriptsize Coefficient arithmetic based DCT  & \scriptsize LLM algorithm & \scriptsize ACT, null mean & \scriptsize ACT, null mean & \scriptsize ACT, Mertens function & \scriptsize ACT, Mertens function \tabularnewline
    \scriptsize Multipliers & \scriptsize 8 & \scriptsize 0 & \scriptsize 0 & \scriptsize 11 & \scriptsize 0 & \scriptsize 0 & \scriptsize 11 & \scriptsize 11\tabularnewline
    \scriptsize Input word-length & \scriptsize 12 & \scriptsize 9 & \scriptsize 9 & \scriptsize 8 & \scriptsize  8 & \scriptsize 12 & \scriptsize 8 & \scriptsize 12\tabularnewline
    \scriptsize Operating frequency (GHz) & \scriptsize 0.125 (2D) & \scriptsize 1.5 & \scriptsize 0.05 (2D) & \scriptsize 0.00489 (2D) & \scriptsize 1.11 & \scriptsize 1.11   & \scriptsize 0.625 & \scriptsize 0.588 \tabularnewline
    \scriptsize Pixel rate ($\times10^9\mathrm{s}^{-1}$) & \scriptsize 0.125 & \scriptsize 12 & \scriptsize 0.4 & \scriptsize 1.792 & \scriptsize 8.88 & \scriptsize 8.88  & \scriptsize 5.00 & \scriptsize 4.70 \tabularnewline
    \scriptsize Power consumption (mW) & \scriptsize N/A & \scriptsize 210  & \scriptsize 12.45 (1D)  & \scriptsize 6.08 (2D) & \scriptsize 67.31~~~\tnote{$\ddagger$} & \scriptsize 90.32~~~\tnote{$\ddagger$} & \scriptsize 60.34~~~\tnote{$\ddagger$} & \scriptsize 79.29~~~\tnote{$\ddagger$} \tabularnewline
    \scriptsize Normalized power consumption (mW/GHz$\cdot$V$^2$) & \scriptsize N/A & \scriptsize 12.86 & \scriptsize 110.67 & \scriptsize 114.17 & \scriptsize 50.11 & \scriptsize 67.25 & \scriptsize 79.79 & \scriptsize 111.44 \tabularnewline
    \scriptsize Gate count & 30290 & N/A & N/A & N/A & 11491 & 16478 & 17673 & 25197\tabularnewline
   \scriptsize Implementation technology &
    \scriptsize 0.25~$\mu$m CMOS &
    \scriptsize 0.35~$\mu$m CMOS &
    \scriptsize 0.12~$\mu$m CMOS &
    \scriptsize 0.35~$\mu$m CMOS &
    \scriptsize 45~nm CMOS &
    \scriptsize 45~nm CMOS &
    \scriptsize 45~nm CMOS &
    \scriptsize 45~nm CMOS
    \tabularnewline
    \scriptsize Supply voltage (V) &
    \scriptsize 2.5 &
    \scriptsize 3.3 &
    \scriptsize 1.5 &
    \scriptsize 3.3 &
    \scriptsize 1.1 &
    \scriptsize 1.1 &
    \scriptsize 1.1 &
    \scriptsize 1.1
    \tabularnewline
    \hline
  \end{tabular}
  \begin{tablenotes}
  \item \scriptsize [$\ddagger$] Dynamic power.
  \end{tablenotes}
\end{threeparttable}
\end{table*}

\section{Conclusions}
\label{section-conclusions}

The ACT algorithm is suitable for calculating the 8-point DCT coefficients
exactly using only adders and integer constant multiplications,
also with low computational complexity.
ACT architectures for null mean inputs as well as for non-null mean inputs are proposed, implemented and
tested on Xilinx Virtex-6 XC6VLX240T FPGA.
The average percentage error and PSNR were adopted as figures of merit
to assess the measured results.
Results show that even for lower fixed point word-lengths,
the implementations lead to acceptable margins of
error.
The resource utilization for various fixed point implementations indicate a trade-off
between accuracy and device resources (chip area, speed, and power).
It is the first step towards new research on
low power and low complexity
computation of the DCT
by means of the recently proposed ACT.

{\small
\bibliographystyle{IEEEtran}
\bibliography{8pt-act-clean}
}

\end{document}